\documentstyle[preprint,tighten,aps]{revtex}

\newcommand{\resetcounter}{\setcounter{equation}{0}}     
\newcommand{\lagrange}{{\cal L}}

\newcommand{\B}  {\beta}

\begin{document}


\draft
\preprint{UPR-753-T, hep-th/9705150}
\date{November 1997}
\title{Gauge Symmetry Enhancement and N = 2 Supersymmetric \\ 
Quantum Black Holes in Heterotic String Vacua}
\author{Ingo Gaida
}

\address{ Department of Physics and Astronomy,
          University of Pennsylvania \\ 
      Philadelphia, PA 19104-6396, U.S.A.}
\maketitle\

\begin{abstract}
$N=2$ supersymmetric quantum black holes in
the heterotic $S$-$T$-$U$ model are presented.
In particular three classes of axion-free quantum black holes
with half the $N=2$, $D=4$ supersymmetries unbroken are considered.
First, these quantum black holes are investigated at generic points
in moduli space. Then ``linearized'' non-abelian black holes are
investigated representing a subset of non-abelian black hole
solutions at critical points of perturbative gauge symmetry enhancement
in moduli space.
It is shown that the entropy of ``linearized'' non-abelian
black holes can be obtained, starting at non-critical points in moduli space, 
by continuous variation of the moduli and a proper identification of
the non-abelian charges.
\\ 
\end{abstract}



%
%

\pacs{04.65.+e,04.70.Dy,11.25.-w,
Keywords: String Theory, S-T-U Model, $N=2$ Supergravity, Black Holes, 
Gauge Symmetry Enhancement.}

\newpage



\section{Introduction}
\resetcounter

Four-dimensional string models including non-perturbative
excitations provide a possibly consistent description
of all interactions. At low-energies it is convenient
to describe these models using an effective supergravity action for
the light string modes where heavy string modes have been integrated out.
At the level of the effective supergravity action, neglecting non-perturbative
effects, the vacuum expectation value of the so-called moduli, i.e. massless
scalar fields with flat potentials, parametrize different string vacua.
The corresponding moduli spaces of four dimensional 
effective string models have a very rich structure. First, they
give rise to duality symmetries.
Second, there exist certain critical points/lines in moduli space giving
rise, for instance, to the stringy version of the Higgs effect 
\cite{SHE,CLM_2}. At these critical points in moduli space a finite number of
additional massless states may appear in the string spectrum
giving rise to gauge symmetry enhancement. In this context the role of
the Higgs field is taken by the moduli. The corresponding one-loop
running coupling constant encounters a logarithmic singularity parametrized
by the Higgs field \cite{CLM_2}. 
This logarithmic singularity takes the threshold
effects of massive modes, becoming massless at the critical points,
into account \cite{CLM_2,Coupling,DKLL,AFGNT,quantum}. 
%

In the context of string theory
it has been shown in \cite{la/wi} that four-dimensional
non-rotating black hole solutions in the BPS limit depend classically
only on the bare quantized charges on the horizon. Thus, the black
hole solutions in the BPS limit are independent of the values of the
moduli at spatial infinity.  In \cite{FerKal1} it has been shown how
one can understand this result from a supersymmetric point of view: On
the horizon the central charge of the extended supersymmetry algebra
acquires a minimal value and thus the extremization of the central
charge provides the specific moduli values on the horizon
\cite{FerKal1,FerGiKal}. 

Although the BPS limit of black hole solutions in four dimensions with
$N \geq 4$ is by now well understood \cite{D4N4Entr}, new features of
black hole physics arise in four-dimensional $N=2$ string theory.  In
particular there exists a large number of different $N=2$ string vacua
so that the extreme black hole solutions depend on the specific
details of the particular $N=2$ string model.

If one considers, for example, 
four-dimensional $N=2$ heterotic string compactifications
on $K3 \times T_{2}$ with $N_{V}+1$ vector multiplets (including the
graviphoton), the classical prepotential is completely universal and
corresponds to a scalar non-linear $\sigma$-model based on the coset space
$\frac{SU(1,1)}{U(1)} \otimes \frac{SO(2,N_{V}-1)}{SO(2) \times SO(N_{V}-1)}$.
However, since
in heterotic $N=2$ string compactifications the dilaton can
be described by a vector multiplet, the heterotic prepotential
receives perturbative quantum corrections at the one-loop level
\cite{DKLL,AFGNT}; in addition there are non-perturbative contributions.

In \cite{BCDKLM,Rey} it has been shown that string loops affect the entropy
of $N=2$ heterotic quantum black holes only through a perturbative
modification of the string coupling. Thus, near critical points 
of perturbative gauge symmetry enhancement in moduli 
space the entropy of $N=2$ supersymmetric quantum black holes
receives logarithmic quantum corrections 
\cite{be/ga,CBG}. In this paper we will also
be concerned with black holes at these critical points themselves.
More results on black hole solutions in $N=2$ supersymmetric vacua are given 
e.g. in \cite{sabra,luest_sabra}

The paper is organized as follows: In section two
we will briefly introduce $N=2$ supergravity, special geometry and
the Bekenstein-Hawking entropy in terms of the $N=2$ prepotential.
In section three we introduce the heterotic $S$-$T$-$U$ model.
Then, in section four, we discuss abelian axion-free quantum black
holes in the $S$-$T$-$U$ model. In section five we investigate
non-abelian quantum black holes associated with critical points of
perturbative gauge symmetry enhancement in moduli space.
Finally we summarize the results and in the appendix
we review, for the sake of completeness, the classical gauge symmetry
enhancement in the $S$-$T$-$U$ model.
 

\section{N = 2 Supergravity and Special Geometry}
\resetcounter
The vector couplings of local N = 2 supersymmetric Yang-Mills theory
are encoded in the holomorphic function $F(X)$, where the $X^{I}$
($I=0 \ldots N_{V}$) denote the complex scalar fields of the vector
supermultiplets. Here $N_{V}$ counts the number of physical scalars, 
and $I$ counts the number of physical vectors. 
The {\em special geometry} \cite{sp}
of this theory can be defined in terms of a symplectic section $V$.
This is a $(2N_{V}+2)$-dimensional complex symplectic vector given by 
$V^T = (X^{I},F_{J})$
with periods $F_{J}= \partial F/\partial X^{J}$. The $N_{V}$ physical scalars
parametrize a $N_{V}$ dimensional complex hypersurface, defined by the
condition that the section satifies a symplectic constraint:
\begin{eqnarray}
i \ 
\left ( 
 \bar X^I F_I - \bar F_I X^I
\right ) &=& 1
\end{eqnarray}
This hypersurface can be described in terms of a complex projective space
with coordinates $z^{A}$ ($A=1, \ldots N_{V}$), if the complex coordinates
are proportional to some holomorphic sections $X^{I}(z)$ of the 
complex projective space:
$X^{I} = e^{K(z,\bar z)/2} X^{I}(z)$
with
\begin{eqnarray}
K(z,\bar z) &=& - \mbox{log} 
\left (
i \bar X^{I}(\bar z) F_{I}( X^{I}(z) ) 
- i X^{I}(z) \bar F_{I} (\bar X^{I}(z)) 
\right ). 
\end{eqnarray}
Moreover one can introduce {\em special coordinates}
$X^{0}(z) = 1$ and $X^{A}(z)= z^{A}$.
In this special coordinates the K\"ahler potential is
\begin{eqnarray}
K(z,\bar z) &=& - \mbox{log} 
\left ( 
2 ( { \cal F} + { \bar{\cal F}} )
- ( z^{A} - \bar z^{A} )( {\cal F}_{A} + { \bar{\cal F}}_{A})
\right ) 
\end{eqnarray}
with $ {\cal F}(z) = i (X^{0})^{-2} F(X)$.

The mass of any physical state is given by the central charge
in the BPS limit:
\begin{eqnarray}
M_{BPS}^{2} &=& |Z|^{2} \ = \ e^{K(z,\bar z)} |q_{I}X^{I}-p^{I}F_{I}|^{2} 
\end{eqnarray}
In the BPS limit the entropy of a black hole is also given by 
the central charge, if the central charge has been extremized
with respect to the moduli ($\partial_{A}|Z| = 0$)\cite{FerKal1}. 
Thus, the moduli take their fixed values at the horizon of the
BPS saturated black hole. This extremization problem is equivalent
to the algebraic solution of the following $2N_{V}+2$ 
``stabilization equations''
\begin{eqnarray}
\label{constraint}
\bar Z V \ - \ Z \bar V &=& i Q
\end{eqnarray}
with the symplectic magnetic/electric charge vector
$Q^{T}=(p^{I},q_{J})$. To solve these equations it is convenient
to go to the so-called Y-basis \cite{BCDKLM}. The corresponding
Y-coordinates are defined as $Y^{I} = \bar Z X^{I}$.
Hence the  $2N_{V}+2$ stabilization equations in the Y-basis are given by
\begin{eqnarray}
Y^{I}- \bar Y^{I} &=& i p^{I},
\hspace{2cm}
F_{I}- \bar F_{I} \ = \ i q_{I}
\end{eqnarray}
and the 
Bekenstein-Hawking entropy in the Y-basis reads
\begin{eqnarray}
S_{BH} &=& i \pi \ \left 
( \bar Y^{I}  F_{I}( Y^{I}) 
-  Y^{I} \bar F_{I} (\bar Y^{I}) 
\right )_{| {\rm fix}} \ = \  \pi \ |Z|_{|{\rm fix}}^{2}.
\end{eqnarray}
Note that the special projective coordinates are invariant under
this change of basis. 


\section{The Perturbative Heterotic S-T-U Model}
\resetcounter

Compactifying the $D=10$ effective heterotic string theory 
on $K3 \times T_{2}$ one can construct the $D=4$, $N=2$
$S$-$T$-$U$ model \cite{ka,theisen}. This model has
244 hypermultiplets, which we will ignore in the following.
Moreover it contains three vector moduli $S$, $T$ and $U$, where $S$ denotes
the heterotic dilaton and $T,U$ the $T_{2}$-moduli.
This model exhibits a non-perturbative symmetry ({\em exchange symmetry}) 
which exchanges the dilaton $S$ with one of the 
two vector moduli $T$ or $U$ \cite{ca/cu,theisen}.  
Moreover the model is invariant under
{\em mirror symmetry} which exchanges $T$ and $U$.

In special projective coordinates the prepotential reads
\begin{eqnarray}
 {\cal F}(S,T,U)  &=& -STU \ + \ h(T,U)
\end{eqnarray}
with 
\begin{eqnarray}
 S &=& -i z^{1},    \hspace{2cm}  
 T \ = \ -i z^{2},  \hspace{2cm}  
U  \ = \ -i z^{3}.
\end{eqnarray}
Here $h(T,U)$ denotes the perturbative quantum corrections.
Then the prepotential in the Y-basis is 
$ F(Y)  = -i (Y^{0})^{2}  {\cal F}(S,T,U)$
with periods  
\begin{eqnarray}
 F_{0}  &=& i Y^{0}  
\left [
-STU  - 2 h + T h_{T}   + U h_{U} 
\right ],
\nonumber\\
 F_{1}  &=& Y^{0}  TU ,
\nonumber\\
 F_{2}  &=& Y^{0}  
\left [ SU  -  h_{T}  \right ],
\nonumber\\
 F_{3}  &=& Y^{0}  
\left [ ST  -  h_{U}  \right ].
\end{eqnarray}
The K\"ahler potential in special coordinates reads
\begin{eqnarray}
K(S,T,U) &=& 
- \mbox{log} ( S + \bar S + V_{GS} )
- \mbox{log} ( T + \bar T )
- \mbox{log} ( U + \bar U ) 
\end{eqnarray}
with 
\begin{eqnarray}
V_{GS}(T,U) &=& 
\frac{ 2( h + \bar h ) - (T + \bar T)(h_{T} + \bar h_{\bar T})  
 - (U + \bar U)(h_{U} + \bar h_{\bar U})}
{ (T + \bar T) (U + \bar U)},
\end{eqnarray}
Here $V_{GS}(T,U)$ denotes the Green-Schwarz term \cite{DKLL},
which yields the true
perturbative target-space duality invariant string coupling
\begin{eqnarray}
{8 \pi\over g^2_{\rm pert}}= S + \bar S+V_{GS}(T,U) 
\;.\label{oneloopc}
\end{eqnarray}
The corresponding semiclassical quantum corrections have been 
determined in \cite{harvey,ca/cu} and read
in the fundamental Weyl chamber Re $T>$ Re $U$
\begin{eqnarray}
h(T,U) &=& - \frac{1}{3} \ U^{3} - c 
 -  \frac{1}{4 \pi^{3}} \sum_{k,l \geq 0} c_{n}(4kl) \ 
    Li_{3} \left ( e^{-2\pi (kT+lU)} \right )
\end{eqnarray}
with $c=\frac{\chi\zeta(3)}{2(2\pi)^4}$ where $\chi=-480$ denotes
the Euler number.
In the semiclassical limit the exchange symmetry is broken, but
the mirror symmetry is still valid.
The singularities of the 
semiclassical prepotential at $T=U \neq 1$, $T=U=1$ and 
$T=U=e^{i\pi/6}$ reflect the perturbative gauge symmetry
enhancement of $U(1)^{2}$ to $SU(2) \times U(1)$, $SU(2)^{2}$ and
$SU(3)$ respectively (see appendix).   
Near these critical points in moduli space the quantum corrections
take the specific form \cite{DKLL,quantum} 
\begin{eqnarray}
h(T,U) &=& \frac{1}{\pi} \ (T-U)^{2} \ \mbox{log} (T-U) + \Delta (T,U)
\nonumber\\
h(T,U) &=& \frac{1}{\pi} \ (T-1) \ \mbox{log} (T-1)^{2} + 
\Delta^{\prime} (T,U)
\nonumber\\ 
h(T,U) &=& \frac{1}{\pi} \ (T-\rho) \ \mbox{log} (T-\rho)^{3} + 
\Delta^{\prime\prime} (T,U).
\end{eqnarray}
Here, the functions $\Delta(T,U)$ are finite and single valued at the
critical points. In the large moduli limit $S,T,U \rightarrow \infty$
(Re $S>$ Re $T>$ Re $U$) the semiclassical quantum corrections reduce to
\begin{eqnarray}
h(T,U) &=& - \frac{1}{3} \ U^{3} - c. 
\end{eqnarray}  
In this particular limit in moduli space the exchange symmetry
is restored perturbatively but mirror symmetry is broken. 
Moreover,
the quantum corrected $S$-$T$-$U$ model
is  dual to the type II model described by 
the elliptically fibered CY space $WP_{1,1,2,8,12}(24)$. 
In the large moduli limit this duality becomes manifest
if one takes for the three K\"ahler class moduli \cite{theisen}
\begin{eqnarray}
\label{type2moduli}
 t_{1}  &=& U, 
\hspace{2cm}
 t_{2}  \ = \ S - T,
\hspace{2cm}
 t_{3}  \ = \ T - U. 
\end{eqnarray}
Ignoring constant contributions the dual prepotential on the 
type II side reads ($t_{2}>t_{3}>0$)
\begin{eqnarray}
\label{type2pre}
 {\cal F}^{0}(t_{1},t_{2},t_{3})  &=& 
- \frac{4}{3} t_{1}^{3} - t_{1}^{2} ( t_{2} + 2 t_{3}) - t_{1} t_{2} t_{3}
- t_{1} t_{3}^{2}.
\end{eqnarray}
In the classical limit ($h=0$) the exchange and mirror symmetry
are both restored and combine together to the classical 
triality symmetry \cite{BKRSW,CLM}.

\section{Abelian Quantum Black Holes in the S-T-U Model}
\resetcounter

In this section we will consider axion-free quantum black holes
in the $S$-$T$-$U$ model, only. This restriction implies
\begin{eqnarray}
 z^{A} (2Y^{0}-ip^{0}) &=& i p^{A}.
\end{eqnarray}
Moreover we will work in a definite 
Weyl chamber. Thus, the perturbative
quantum corrections $h(T,U)$ are real.
This implies that our results are not manifest target-space
duality invariant \cite{CLM,BCDKLM,CBG}. 
On the other hand, all the results we will obtain can be given
in a manifest target-space duality invariant form following
\cite{CBG}. 


\subsection{Axion-free Quantum Black Holes in the S-T-U Model}


In particular we will discuss three different dyonic axion-free
black hole configurations (i)-(iii):

\subsubsection{First class of dyonic axion-free quantum black holes}

In this case we take $Y^{0}+\bar Y^{0}=\lambda \neq 0$.
Then the first set of stabilization equations yields 
\begin{eqnarray}
 S  &=&  \frac{p^{1}}{\lambda}, \hspace{2cm}  
 T \ = \ \frac{p^{2}}{\lambda}, \hspace{2cm}  
 U \ = \ \frac{p^{3}}{\lambda},
\end{eqnarray}
and from the second set we obtain
\begin{eqnarray}
h &=& \frac{1}{2p^{0}\lambda} (p^{1}q_{1}-p^{2}q_{2}-p^{3}q_{3}-p^{0}q_{0}), 
\hspace{1cm}
\lambda \ = \ \pm \sqrt{\frac{p^{0}p^{2}p^{3}}{q_{1}}}, 
\nonumber\\
h_{T} &=& \frac{p^{1}q_{1}}{p^{0}p^{2}} - \frac{q_{2}}{p^{0}}, 
\hspace{1cm}
h_{U} \ = \ \frac{p^{1}q_{1}}{p^{0}p^{3}} - \frac{q_{3}}{p^{0}}.
\end{eqnarray}
In the classical limit this yields $STU=-q_{0}/\lambda$ and the 
classical charge constraints
\begin{eqnarray}
 p^{1}q_{1} &=& p^{2}q_{2} \ = \ p^{3}q_{3} \ = \ -p^{0}q_{0}.  
\end{eqnarray}
The classical entropy is given by
\begin{eqnarray}
 S_{BH}^{class}  &=& 2\pi \ 
                  \frac{\lambda^{2}+(p^{0})^{2}}{p^{0}\lambda} \
                   p^{1}q_{1}. 
\end{eqnarray}
The restrictions from the stabilization equations are such that
the quantum corrections of the entropy are of a restricted form, i.e.
the charges have to obey constraints.
In the case of perturbative quantum corrections the charges obey the 
general semiclassical constraint
\begin{eqnarray}
 p^{1}q_{1}  &=&  p^{2}q_{2} + p^{0} p^{2} h_{T}
\ = \
 p^{3}q_{3} + p^{0} p^{3} h_{U}
\ = \ 
 p^{2}q_{2} +  p^{3}q_{3} + p^{0} q_{0} + 2 p^{0} \lambda h
\end{eqnarray}
and the entropy reads
\begin{eqnarray}
\label{qentropy1}
 S_{BH}  &=& 
\label{case_1}
\frac{\pi}{2} \ \frac{\lambda^{2}+(p^{0})^{2}}{p^{0}\lambda} \
            \left (
p^{1}q_{1} +  p^{2}q_{2} +  p^{3}q_{3} - p^{0}q_{0}
            \right )
\end{eqnarray}
This entropy includes all perturbative quantum corrections.
The parameter $\lambda$ can be determined because
the perturbative quantum corrections are independent of the
dilaton \cite{DKLL,AFGNT}. On the other hand, 
including non-perturbative quantum corrections
one finds for this black hole configuration the same entropy
(\ref{case_1}), but the parameter $\lambda$ remains an
undetermined parameter.
\subsubsection{Second class of dyonic axion-free quantum black holes}

For this configuration we take  
$Y^{0}+\bar Y^{0}=0$. Then we obtain from the first set of stabilization
equations
$Y^{0} = \frac{i}{2} p^{0}$ and $ p^{A}= 0$.
The second set yields
\begin{eqnarray}
 q_{0}  &=& 0                         , \hspace{0,5cm}
    TU \ = \ \frac{q_{1}}{p^{0}}      , \hspace{0,5cm}
 h_{T} \ = \ SU - \frac{q_{2}}{p^{0}} , \hspace{0,5cm}
 h_{U} \ = \ ST - \frac{q_{3}}{p^{0}}. 
\end{eqnarray}
In the classical limit one finds for the fixed values of
the moduli on the horizon
\begin{eqnarray}
 S  &=&  \sqrt{ \frac{q_{2}q_{3}}{q_{1}p^{0}} }, \hspace{1cm}  
 T \ = \ \sqrt{ \frac{q_{1}q_{3}}{q_{2}p^{0}} }, \hspace{1cm}  
 U \ = \ \sqrt{ \frac{q_{1}q_{2}}{q_{3}p^{0}} }
\end{eqnarray}
and the classical entropy reads
\begin{eqnarray}
 S_{BH}^{class}  &=& 2 \pi \ \sqrt{p^{0}q_{1}q_{2}q_{3}}. 
\end{eqnarray}
In the case of quantum corrections, on the other hand, 
the entropy is in general
\begin{eqnarray}
 S_{BH}  &=& 
\pi p^{0} \ \left ( q_{2}T + q_{3}U + p^{0}h \right )_{| \rm fix} .
\end{eqnarray}
Here, $T$ and $U$ take their fixed values on the horizon.
To find these fixed values without further restrictions on the
charges or limits in moduli space seems to be difficult at this point.
\subsubsection{Third class of dyonic axion-free quantum black holes}
For this configuration we take 
$Y^{0}-\bar Y^{0}=0$. Thus, the first set of stabilization
equations yields $Y^{A}=\frac{i}{2}p^{A}$ and hence
\begin{eqnarray}
 S  &=&  \frac{p^{1}}{2Y^{0}},  \hspace{1cm}  
 T \ = \ \frac{p^{2}}{2Y^{0}},  \hspace{1cm}  
 U \ = \ \frac{p^{3}}{2Y^{0}}.
\end{eqnarray}
From the second set one obtains
\begin{eqnarray}
 q_{A}  &=& 0, \hspace{2cm}
q_{0} \ = \ 2Y^{0} [-STU-2h+Th_{T}+Uh_{U}].
\end{eqnarray}
In the classical limit, with $q_{0}<0$, the fixed values
of the moduli on the horizon are
\begin{eqnarray}
 Y^{0} &=& \frac{1}{2} \sqrt{\frac{p^{1}p^{2}p^{3}}{|q_{0}|}}, \hspace{0,5cm}
 S \ = \  \sqrt{\frac{p^{1}|q_{0}|}{p^{2}p^{3}}}, \hspace{0,5cm}  
 T \ = \ \sqrt{ \frac{p^{2}|q_{0}|}{p^{1}p^{3}}}, \hspace{0,5cm}  
 U \ = \ \sqrt{ \frac{p^{3}|q_{0}|}{p^{1}p^{2}}}
\end{eqnarray}
and the corresponding classical entropy reads
\begin{eqnarray}
 S_{BH}^{class}  &=& 2 \pi \ \sqrt{|q_{0}|p^{1}p^{2}p^{3}}. 
\end{eqnarray}
Moreover, in the case of quantum corrections the entropy is in general
\begin{eqnarray}
 S_{BH}  &=& 4 \pi \ (Y^{0})^{2} 
\left (
2STU + h - h_{T}T - h_{U} U
\right )_{| \rm fix}
\end{eqnarray}
with
\begin{eqnarray}
 Y^{0}  &=& - \frac{q_{0}}{2} 
\left (
STU + 2h - h_{T}T - h_{U} U
\right )^{-1}_{| \rm fix}.
\end{eqnarray}
Near the critical point of perturbative gauge symmetry enhancement 
$T=U \neq 1$ one finds that $Y^{0}$ is single valued
\begin{eqnarray}
 Y^{0}  &=& - \frac{q_{0}}{2} 
\left (
STU - \frac{1}{\pi}(T-U)^{2}+2\Delta - \Delta_{T}T - \Delta_{U} U
\right )_{| \rm fix}^{-1}.
\end{eqnarray}
The corresponding quantum corrected entropy reads
\begin{eqnarray}
 S_{BH}  &=& 4\pi \ (Y^{0})^{2} 
\left (
2STU - \frac{1}{\pi} (T-U)^{2} [\mbox{log} (T-U) +1] + \Delta 
- \Delta_{T}T - \Delta_{U} U
\right )_{| \rm fix}.
\nonumber\\
\end{eqnarray}
Thus, the entropy of the quantum black hole receives 
logarithmic and polynomial quantum corrections 
due to one-loop effects in the effective
action of the $S$-$T$-$U$ model. 

Moreover, at the microscopic level
the logarithmic quantum corrections
are a possible origin of subleading terms in the degeneracy of
an underlying (unknown) quantum theory \cite{be/ga}.
If we consider, for example, the dyonic case at hand
and omit polynomial quantum corrections encoded in $\Delta(T,U)$ we find 
\begin{eqnarray}
 Y^{0}  &=& \frac{1}{4\pi} \frac{(p^{2}-p^{3})^{2}}{q_{0}}
+ \sqrt{
 \left ( \frac{1}{4\pi} \frac{(p^{2}-p^{3})^{2}}{q_{0}} \right )^{2}
- \frac{p^{1}p^{2}p^{3}}{4q_{0}}
}.
\end{eqnarray}
Now we consider the large $q_0$ limit ($q_{0}<0, p^{A}>0$) and find that
$Y^{0}$ is given by its classical value. Thus, the quantum corrected
entropy in this limit reads 
\begin{eqnarray}
\label{largeN}
 S_{BH}  &=& 2 \pi \  \sqrt{|q_0| p^{1}p^{2}p^{3}} 
 - \frac{1}{2} (p^{2}-p^{3})^{2} \ \mbox{log} \ 
\left (
 \frac{ (p^{2}-p^{3})^{2}|q_0|}{p^1 p^2 p^3} 
\right ) + \cdots
\end{eqnarray}
Here the dots stand for polynomial quantum corrections encoded in
$\Delta(T,U)$. Eq. (\ref{largeN}) is the perturbative corrected black hole
entropy in a special region of moduli space and represents the
entropy of an abelian black hole. However, if we consider the
limit $T \rightarrow U$ ($p^2 \rightarrow p^3$) in moduli space,
the $U(1)$ gauge symmetry becomes enhanced to $SU(2)$ and the effective action 
changes. 
This is a pure stringy effect and reflects the fact that at
the line $T=U$ additional states become massless (see appendix)
giving rise to a perturbative gauge symmetry enhancement. 
The quantum corrected entropy approaches this line in moduli space
smooth, because the perturbative quantum corrections already take
this light states, becoming massless at $T=U$, into account \cite{quantum}.
Thus, approaching the line of perturbative gauge symmetry enhancement,
the logarithmic quantum corrections in (\ref{largeN}) vanish. However,   
on the line of perturbative gauge symmetry enhancement
(\ref{largeN}) is, first of all, not the correct entropy, since
the effective action that has been used to calculate the entropy, is not
correct anymore. Instead the effective action has a non-abelian
$SU(2)$ sector on the line $T=U$. Of course, if the black hole
solution breaks this non-abelian gauge group explicitly down
to $U(1)$, we can take the limit $T=U$ explicitly.


\subsection{Dual Quantum Black Hole Pairs in the S-T-U Model }


In this subsection we will consider the
large moduli limit of the $S$-$T$-$U$ model.
In this limit the $S$-$T$-$U$ model is dual to 
certain CY compactifications of type II string models. 
Here we consider, as an example, the CY space described 
by $WP_{1,1,2,8,12}(24)$.
An analogous discussion for other CY spaces using,
for instance, the results of \cite{theisen} is straightforward.

Classical $N=2$ supersymmetric black holes in the context of Calabi-Yau
compactifications have been already extensively discussed in
\cite{BKRSW,BCDKLM,be/mo,shmakova}. Nevertheless it is instructive, 
at this point of our discussion, to consider these heterotic 
quantum black holes explicitly.

Let us recall that in the large moduli limit the mirror symmetry 
is broken, the exchange symmetry, on the other hand, is valid. Thus, the
corresponding black hole solutions are in the type II and
the heterotic description identical and have, in addition,
a {\em perturbative} exchange symmetry. In the following
we will consider again the three different dyonic classes 
(i) - (iii) in the large moduli limit:  
\subsubsection{First class of dual quantum black hole pairs}
In this particular configuration the fields on the heterotic
side are already determined
in terms of the charges.
The charges obey the constraints
\begin{eqnarray}
\label{lmc1}
 p^{1}q_{1} &=& p^{2}q_{2} 
\ = \ p^{3}q_{3} - (p^{3})^{2} \frac{q_{1}}{p^{2}} 
\ = \  p^{2}q_{2} + p^{3}q_{3} +  p^{0}q_{0} 
       - \frac{2}{3} (p^{3})^{2} \frac{q_{1}}{p^{2}}  
\end{eqnarray}
and the quantum corrected entropy is (\ref{qentropy1}) obeying
(\ref{lmc1}). In addition the exchange symmetry
exchanges $p^{1} \leftrightarrow p^{2}$ and
$q_{1} \leftrightarrow q_{2}$ leaving invariant the constraints
on the charges and the parameter $\lambda$.
\subsubsection{Second class of dual quantum black hole pairs}
In this particular configuration
one obtains for the fixed values of the heterotic moduli
\begin{eqnarray}
 S  &=&  \sqrt{\frac{ 2(q_{2})^{2}}{\B q_{3} p^{0}} }, \hspace{1cm}  
 T \ = \ \sqrt{\frac{ 2(q_{1})^{2}}{\B q_{3} p^{0}} }, \hspace{1cm}  
 U \ = \ \sqrt{\frac{ \B  q_{3}}{2p^{0}} } 
\end{eqnarray}
with $\B = 1 - \sqrt{1 - 4 \frac{q_{1}q_{2}}{(q_{3})^{2}} }$.
Thus, the exchange symmetry exchanges $p^{1} \leftrightarrow p^{2}$. 
The corresponding quantum corrected entropy reads
\begin{eqnarray}
 S_{BH}  &=& 
\pi \ p^0 q_1 q_2 q_3 \sqrt{\frac{2 p^0}{\B q_3^3}}
+ \pi \sqrt{\frac{p^0 \B q_3^3}{2}}
+ \pi (p^0)^2 h(T,U)_{| \rm fix} 
\end{eqnarray} 
If we consider the limit of small electric charges 
($q_A \ll 1$) and large magnetic charge $p^0 \gg 1$, then
the leading contribution to the entropy has its origin in
the constant part of the perturbative quantum corrections.
In this particular limit we find
\begin{eqnarray}
 S_{BH}  &=& \pi (p^0)^2 h(0,0) + \cdots
\ = \ - \frac{\chi}{4 (2\pi)^3} (p^0)^2 \zeta(3) + \cdots
\end{eqnarray}
Thus, we recover the result of \cite{CBG} that in a particular limit
in moduli space the leading contribution to the entropy is proportional
to $\zeta(3)$. Note that in this particular context the leading contribution
is finite.
\subsubsection{Third class of dual quantum black hole pairs}
For this configuration we obtain, first of all,
\begin{eqnarray}
 Y^{0}  &=&  \frac{1}{2} 
\sqrt{\frac{p^{1}p^{2}p^{3} + (p^{3})^{3}}{3 |q_{0}|}}.
\end{eqnarray}
Thus, the fixed values of the heterotic moduli are given by
\begin{eqnarray}
 S  &=&  \sqrt{\frac{(p^{1})^{2}|q_{0}|}{p^{1}p^{2}p^{3}+(p^{3})^{3}/3}}, 
\hspace{1cm}  
 T \ = \  \sqrt{\frac{(p^{2})^{2}|q_{0}|}{p^{1}p^{2}p^{3}+(p^{3})^{3}/3}}, 
\hspace{1cm}  
 U \ = \  \sqrt{\frac{p^{3}|q_{0}|}{p^{1}p^{2}+ (p^{3})^{2}/3}}.
\nonumber\\
\end{eqnarray}
Again the exchange symmetry exchanges $p^{1} \leftrightarrow p^{2}$ and 
the corresponding quantum corrected entropy reads
\begin{eqnarray}
 S_{BH}  &=& 2\pi \ 
\sqrt{|q_{0}|p^{1}p^{2}p^{3} + \frac{|q_{0}|}{3}(p^{3})^{3}}. 
\end{eqnarray}

\section{Non-abelian Quantum Black Holes in the S-T-U Model}
\resetcounter

In the previous section we have considered quantum black holes
at generic points in moduli space, i.e. the effective four-dimensional
supergravity action had only abelian gauge groups.
In this section we will discuss critical points in moduli space,
where the effective action also has a non-abelian gauge sector.
Thus, the corresponding black hole solutions can be non-abelian. 

There are two classes of non-abelian black holes: i) linearized
black hole solutions \cite{Yasskin} and ii) non-linear black
hole solutions \cite{N_BH}. 

The linearized non-abelian black hole solutions i) are related to 
abelian black hole solutions by construction. 
In the following we will discuss these linearized non-abelian
black hole solutions, only. Clearly these solutions represent only a
certain subset of solutions of the equations of motion.
Moreover, the linearized non-abelian black hole solutions of \cite{Yasskin}
have been studied in a pure Yang-Mills context. In order to study
linearized non-abelian black holes in string vacua, one must 
reformulate the theorem given in \cite{Yasskin}.
\subsection{Linearized solution theorem in string theory}
The ``linearized solution theorem'' of \cite{Yasskin} in the
context of string theory can be formulated as follows
\\
\\
{\underline{Theorem:}} 
{\em 
Let ${\cal G}$ be a N-parameter Lie-group with an invariant
metric $\gamma_{ab}$ ($a,b=1 \ldots N$). Then for every solution of the
field-dependent (source-free) coupled Einstein-Maxwell equations there
is a $(N-1)$ parameter set of solutions of the
field-dependent coupled Einstein-massless-Yang-Mills equations for
the gauge group ${\cal G}$.
}
\\
\\
{\underline{Proof:}} The field-dependent coupled Einstein-Maxwell
Lagrangian is in general of the form 
\begin{eqnarray}
\label{EM_1}
 e^{-1} \lagrange_{EM}  &=&  R + f(\phi) F^2 + g(\phi) F \tilde F
\end{eqnarray}
in the Einstein frame. Here $f(\phi),g(\phi)$ are arbitrary
field-dependent functions with $\phi_i$ ($i=1, \ldots$), 
At tree level in heterotic string vacua, for instance,
$f(\phi)$ is given by the dilaton and $g(\phi)$ by
the model-independent axion. The corresponding solution of the
field-dependent coupled
Einstein-Maxwell equations of motion of (\ref{EM_1}) are given by 
$g_{\mu\nu}^0,A_\mu^0,\phi_i^0$. Then the solution of the
field-dependent coupled Einstein-Yang-Mills equation for 
gauge group ${\cal G}$ with metric $\gamma_{ab}$ 
are given by $g_{\mu\nu},A_\mu^a,\phi_i$ with
\begin{eqnarray}
\label{c_1}
  g_{\mu\nu} &\equiv& g_{\mu\nu}^0, \hspace{1cm}
  A_\mu^a \ = \ \B^a A_\mu^0, \hspace{1cm}
  \phi_i \ \equiv \ \phi_i^0.
\end{eqnarray}
The N-parameters $\B^a$ are subject to the constraint
\begin{eqnarray}
\label{c_2}
 \langle \B | \B \rangle &=& 1.
\end{eqnarray}  
Here, we have defined a scalar product 
$\langle X | X \rangle=X^a \gamma_{ab} X^b$. The Lagrangian of
the field-dependent coupled Einstein-Yang-Mills model
is given by
\begin{eqnarray}
\label{EYM_1}
 e^{-1}\lagrange_{EYM}  &=&  R + f(\phi) \mbox{tr} F^2 + 
                      g(\phi)  \mbox{tr} F \tilde F
\end{eqnarray} 
with $ \mbox{tr} F^2= F_{\mu\nu}^{ \ \ a} \gamma_{ab} F^{\mu\nu b}$
and an analogous expression for $ \mbox{tr} F \tilde F$.
Note that the field-dependent coupling functions $f(\phi),g(\phi)$
have to be the {\em same} as in the pure abelian case. 
Using now (\ref{c_1})
one obtains\footnote{Note that the generators $T_a \in {\cal G}$ obey 
$[T_a,T_b]=f_{ab}^{ \ \ c} T_c$ with antisymmetric structure constants
$f_{abc}$.}
$F_{\mu\nu}^{ \ \ a}=
\partial_\mu A_\nu^a -\partial_\nu A_\mu^a=\B^a  F_{\mu\nu}^0$
and therefore $\lagrange_{EYM}=\lagrange_{EM}$. Since the
same action has the same solutions of the equations of motion, the proof is
complete.

Note that the spacetime symmetry of the abelian and the non-abelian
solution are the same. The non-abelian solution, however, depends on
$N-1$ independent parameters. These parameters
are related to the non-abelian charges: 
The electric and magnetic charges of the non-abelian
solution, given by $F_{0r}^{ \ a} \sim \frac{Q^a}{r^2}$ 
and $\tilde F_{0r}^{ \ a} \sim \frac{P^a}{r^2}$ for large $r$,
are
\begin{eqnarray}
 Q^a &=& \B^a \ Q^0, \hspace{2cm}  P^a \ = \ \B^a \ P^0.  
\end{eqnarray}
Here ($Q^0,P^0$) denote the electric and magnetic charge of the
corresponding abelian solution respectively. 
Using (\ref{c_2}) we find that
the charges have to obey the charge constraints
\begin{eqnarray}
\label{ch1}
  \langle Q | Q \rangle &=& (Q^0)^2, \hspace{2cm}
  \langle P | P \rangle \ = \ (P^0)^2
\end{eqnarray}
and Dirac's quantization condition
\begin{eqnarray}
\label{ch2}
  \langle P | Q \rangle &=& Q^0 P^0  \ = \ 2 \pi n. 
\end{eqnarray} 
The difference between the abelian and the non-abelian solution is
that in the non-abelian case the solution depends on $N-1$ independent
non-abelian charges. The particular 
dependence is constrained by (\ref{ch1}) and (\ref{ch2}). 
These linearized non-abelian solutions contain the
case of a pure abelian solution, which is reached if
$\B^1 = 1$ and $\B^n = 0$ with $n=2, \ldots, N-1$. 

\subsection{Application to quantum black holes} 
From the ``linearized solution theorem'' follows: The
entropy of a given abelian and non-abelian black hole solution
is identical, if they satisfy the ``linearized solution conditions''
(\ref{c_1}) and (\ref{c_2}). Thus, for the case at hand, if we approach
lines/points of gauge symmetry enhancement in moduli space, the
entropy, derived in the abelian effective field theory, is still
valid at the critical points themselves, if the non-abelian black hole
solution satisfies (\ref{c_1}) and (\ref{c_2}). 
One must only identify the
non-abelian charges in terms of the abelian ones. 
To be more concrete, let us consider the classical
dyonic solutions (i)-(iii) at the 
critical line $T=U$. For the first class of
dyonic axion-free black hole solutions
we find $p^2=p^3$ and $q_2=q_3$. Thus, the entropy of the non-abelian
black hole solution reads
\begin{eqnarray}
 S_{BH}^{class}  &=& 2\pi \ 
                  \frac{\lambda^{2}+(p^{0})^{2}}{p^{0}\lambda} \
                   p^{1}q_{1},
\hspace{2cm} 
\lambda = \pm \sqrt{\frac{p^0}{q_1} \langle P | P \rangle}.
\end{eqnarray}
Here $P^a$ ($a=1,2,3$) denote the non-abelian magnetic charges
corresponding to the $SU(2)$ group.
Analogous one finds for the second class (ii) 
$(q_2)^2=(q_3)^2=  \langle Q | Q \rangle $
and for the third class (iii)
$(p^2)^2=(p^3)^2=  \langle P | P \rangle $. 
Finally,
since we can approach the lines/points of perturbative gauge symmetry
enhancement explicitly, if 
the corresponding non-abelian black hole solution
satisfies the ``linearized solution theorem'', we find that quantum corrections
yield polynomial corrections to the entropy
at these critical points, only. 
Note that these points of perturbative
gauge symmetry enhancement are non-critical in the sense that
no phase transitions as discussed in \cite{rahmfeld} occur.
Moreover, the non-abelian black holes
satisfying the ``linearized solution theorem'' do not really
create a so-called ``gauge charge hair'' \cite{Yasskin,N_BH,hair}. This can be
seen already via the charge constraints (\ref{ch1})
and (\ref{ch2}). The linearized black hole
solution does not depend on particular data with respect to the
gauge group other than the charges\footnote{ 
Note that non-extreme black holes in string theory 
depend on the values of the moduli at infinity \cite{youm}.}.
\section{Summary and Conclusion}
$N=2$ supersymmetric quantum black holes in the
heterotic $S$-$T$-$U$ model have been studied. 
First, three classes of axion-free quantum black holes
with half the $N=2$, $D=4$ supersymmetries unbroken have been
investigated. Remarkably, the entropy of one
class of solutions is valid everywhere at generic points in moduli space,
while the entropy of the other two classes depend on the specific 
perturbative quantum
corrections in moduli space. However, these results were, first of all,
only valid at generic points in moduli space, where the corresponding
low-energy effective action contains only abelian gauge groups.
In the second part we considered ``linearized'' non-abelian black holes.
It is shown that the entropy of linearized non-abelian black holes 
at critical points of perturbative gauge symmetry enhancement can be reached, 
starting at non-critical points, 
by continuous variation of the moduli and a proper identification of
the non-abelian charges. These linearized non-abelian black hole solutions
contain the pure abelian black hole solutions valid already 
at generic points in moduli space. On the other hand, the linearized 
non-abelian black hole solutions represent only a subset of solutions in the
general non-abelian $S$-$T$-$U$ model. Thus, it would be very interesting 
to consider also non-linear non-abelian black hole solutions 
(see e.g. \cite{N_BH,volkov}) in the context of the $S$-$T$-$U$ model
at critical points of perturbative gauge symmetry enhancement.   
 \bigskip 

\noindent
{\bf Acknowledgments}  \medskip \newline
I would like to thank T. Mohaupt, M. Cveti{\v{c}},
F. Larsen, G. Lopes-Cardoso and especially K. Behrndt
for helpful discussions.
This work 
~is supported by U.S. DOE Grant Nos. DOE-EY-76-02-3071 and the National 
Science Foundation Career Advancement Award No. PHY95-12732.

\newpage

\section{Appendix}
The heterotic string on $K3 \times T_2$ has two moduli $T$ and $U$
corresponding to the torus $T_2$.
These two scalars are members of two $U(1)$ $N=2$ vector
multiplets at generic points in moduli space.
In addition to the corresponding $U(1)_L \times U(1)_R$ symmetry of $T_2$
the model contains the heterotic dilaton and the graviphoton.
Thus, at generic points in moduli space there is a 
$U(1)_L^2 \times U(1)^2_R$ abelian symmetry. At special 
lines/points in the classical $(T,U)$ moduli space additional 
gauge bosons become massless and the $U(1)^2_L$ becomes enlarged
to a non-abelian gauge symmetry.
First of all there are four inequivalent lines in the
classical moduli space where two charged gauge bosons become massless 
\cite{CLM_2,quantum}: 
Using the mass formula for $N=2$ BPS states one finds 
\begin{eqnarray}
 M_{BPS}^2 &=& e^{K(z,\bar z)} |{\cal M} (S,T,U)|^2
\ \sim \  |{\cal M} (T,U)|^2 \ = \ 
|m_2-i m_1 U + i n_1 T - n_2 T U|^2.
\end{eqnarray}  
Here,
${\cal M}$ denotes the holomorphic mass \cite{hol_mass} and
$m_{1,2}$ [$n_{1,2}$] the momentum [winding] quantum numbers in the
$1,2$ direction of $T_2$. The four critical lines
of perturbative gauge symmetry enhancement can be read off 
straightforward.

\begin{center}
\begin{tabular}{|c|c|c|}
\hline
\ critical lines \ & quantum numbers & \ gauge symmetry \  \\
\hline
$T=U$              & $m_1=n_1=\pm 1, m_2=n_2=0$ & $SU(2)_L \times U(1)_L$ \\
\hline
$T=1/U$            & $m_1=n_1=0 , m_2=n_2=\pm 1$& $SU(2)_L \times U(1)_L$ \\
\hline
$T=U+i$            & $m_1=n_1=m_2=\pm 1 ,n_2=0$ & $SU(2)_L \times U(1)_L$ \\
\hline
$T=\frac{U}{iU+1}$ & $m_1=n_1=-n_2=\pm 1 , m_2=0$&$SU(2)_L \times U(1)_L$ \\
\hline
\end{tabular}
\end{center}

If two [three] critical lines intersect with each other four
[six] additional states, corresponding
to gauge bosons, become massless at the intersection point.
At this points the gauge group is enlarged to $SU(2)^2_L$ 
[$SU(3)_L$].

\begin{center}
\begin{tabular}{|c|c|c|}
\hline
\ critical points \ & quantum numbers & \ gauge symmetry \  \\
\hline
$T=U=1$       & $m_1=n_1=\pm 1, m_2=n_2=\pm 1$ & $SU(2)_L \times SU(2)_L$ \\
\hline
              & $m_2=n_2=0 , m_1=n_1=\pm 1$&    \\
              & $m_2=n_2=1 , m_1=-1 , n_1=0 $&  \\
$T=U=e^{i\pi/6}$& $m_2=n_2=1 , m_1=0, n_1=1$& $SU(3)_L$  \\
              & $m_2=n_2=-1 , m_1=0,n_1=-1$&              \\
              & $m_2=n_2=-1 , m_1=1, n_1=0$&              \\
\hline
\end{tabular}
\end{center}

%
%

\renewcommand{\arraystretch}{1}

\newcommand{\NP}[3]{{ Nucl. Phys.} {\bf #1} {(19#2)} {#3}}
\newcommand{\PL}[3]{{ Phys. Lett.} {\bf #1} {(19#2)} {#3}}
\newcommand{\PRL}[3]{{ Phys. Rev. Lett.} {\bf #1} {(19#2)} {#3}}
\newcommand{\PR}[3]{{ Phys. Rev.} {\bf #1} {(19#2)} {#3}}
\newcommand{\IJ}[3]{{ Int. Jour. Mod. Phys.} {\bf #1} {(19#2)}
  {#3}}
\newcommand{\CMP}[3]{{ Comm. Math. Phys.} {\bf #1} {(19#2)} {#3}}
\newcommand{\PRp} [3]{{ Phys. Rep.} {\bf #1} {(19#2)} {#3}}

\end{document}